\documentclass
[bibnotes,balancelastpage,secnumarabic,prl,onecolumn,hyperref,preprint,titlepage,unsortedaddress,11pt,tightenlines]{revtex4}%
\usepackage{amsfonts}
\usepackage{amsmath}
\usepackage{amssymb}
\usepackage{graphicx}
\usepackage{hyperref}%
\providecommand{\U}[1]{\protect\rule{.1in}{.1in}}

\begin{document}
\preprint{UMTG-27}
\title[Naked Gal]{Galileons and Naked Singularities}
\author{Thomas L. Curtright}
\affiliation{Department of Physics, University of Miami, Coral Gables, FL 33124-8046,
USA\medskip\medskip}

\begin{abstract}
A simple trace-coupled Galileon model is shown to admit\ spherically symmetric
static solutions with naked spacetime curvature singularities.

\end{abstract}
\volumeyear{year}
\volumenumber{number}
\issuenumber{number}
\eid{identifier}
\startpage{1}
\endpage{ }
\maketitle

Galileon theories are a class of models for hypothetical scalar fields whose
Lagrangians involve multilinears of first and second derivatives, but whose
nonlinear field equations are still only second order. \ They may be important
for the description of large-scale features in astrophysics as well as for
elementary particle theory \cite{deRham,Fairlie}. \ Hierarchies of Galileon
Lagrangians were first discussed mathematically in \cite{FairlieEtAl.}. \ The
simplest example involves a single scalar field, $\phi$. \ This Galileon field
is usually coupled to all \emph{other} matter through the trace of the
energy-momentum tensor, $\Theta^{\text{(matter)}}$, and is thus
gravitation-like by virtue of the similarity between this universal coupling
and that of the metric $g_{\mu\nu}$ to $\Theta_{\mu\nu}^{\text{(matter)}}$ in
general relativity. \ In fact, some Galileon models have been obtained from
limits of higher dimensional gravitation theories \cite{DGPetc.}.

In \cite{CF}\ the effects of coupling a Galileon to its own energy-momentum
trace were considered, in the flat spacetime limit. \ Here, general
relativistic effects are taken into consideration and additional features of
this same model are explored in curved spacetime \cite{DeffayetEtAl.}. \ While
this investigation was in progress, I learned of other work \cite{CharmEtAl.}
for a related class of models. \ However, one of the main points of that other
study is at variance with the results to follow, namely, the model discussed
here admits solutions with \emph{naked singularities} when the energy in the
scalar field is finite and not too large, and for which the effective mass of
the system is positive. \ Thus for the simple model at hand there is an open
set of \emph{physically acceptable} scalar field data for which curvature
singularities are \emph{not} hidden inside event horizons \cite{Rama,Rinaldi}.
\ This would seem to have important implications for the cosmic censorship
conjecture \cite{Penrose,Wald,Singh}. \ It is worthwhile to note that, in
general, naked singularities have observable consequences that differ from
those due to black holes \cite{VirbhadraEtAl.}.

The scalar field part of the action in curved space is
\begin{equation}
A=\frac{1}{2}\int g^{\alpha\beta}\phi_{\alpha}\phi_{\beta}\left(  1-\frac
{1}{\sqrt{-g}}~\partial_{\mu}\left(  \sqrt{-g}g^{\mu\nu}\phi_{\nu}\right)
-\frac{1}{2}~g^{\mu\nu}\phi_{\mu}\phi_{\nu}\right)  \sqrt{-g}~d^{4}x\ .
\label{Action}%
\end{equation}
This gives a symmetric energy-momentum tensor $\Theta_{\alpha\beta}$ for
$\phi$ upon variation of the metric. \
\begin{equation}
\delta A=\tfrac{1}{2}\int\sqrt{-g}~\Theta_{\alpha\beta}~\delta g^{\alpha\beta
}~d^{4}x\ ,
\end{equation}%
\begin{align}
\Theta_{\alpha\beta}  &  =\phi_{\alpha}\phi_{\beta}\left(  1-g^{\mu\nu}%
\phi_{\mu}\phi_{\nu}\right)  -\tfrac{1}{2}g_{\alpha\beta}~g^{\mu\nu}\phi_{\mu
}\phi_{\nu}\left(  1-\tfrac{1}{2}g^{\rho\sigma}\phi_{\rho}\phi_{\sigma}\right)
\nonumber\\
&  -\phi_{\alpha}\phi_{\beta}\tfrac{1}{\sqrt{-g}}\partial_{\mu}\left(
\sqrt{-g}g^{\mu\nu}\phi_{\nu}\right)  +\tfrac{1}{2}\partial_{\alpha}\left(
g^{\mu\nu}\phi_{\mu}\phi_{\nu}\right)  \phi_{\beta}+\tfrac{1}{2}%
\partial_{\beta}\left(  g^{\mu\nu}\phi_{\mu}\phi_{\nu}\right)  \phi_{\alpha
}-\tfrac{1}{2}g_{\alpha\beta}\partial_{\rho}\left(  g^{\mu\nu}\phi_{\mu}%
\phi_{\nu}\right)  g^{\rho\sigma}\phi_{\sigma}\ . \label{EnMomTensor}%
\end{align}
It also gives the field equation for $\phi$ upon variation of the scalar
field, $\mathcal{E}\left[  \phi\right]  =0$, where
\begin{gather}
\delta A=-\int\sqrt{-g}~\mathcal{E}\left[  \phi\right]  ~\delta\phi
~d^{4}x\ ,\\
\mathcal{E}\left[  \phi\right]  =\partial_{\alpha}\left[  g^{\alpha\beta}%
\phi_{\beta}\sqrt{-g}-g^{\alpha\beta}\phi_{\beta}~g^{\mu\nu}\phi_{\mu}%
\phi_{\nu}\sqrt{-g}-g^{\alpha\beta}\phi_{\beta}\partial_{\mu}\left(  \sqrt
{-g}g^{\mu\nu}\phi_{\nu}\right)  +\tfrac{1}{2}\sqrt{-g}g^{\alpha\beta}%
\partial_{\beta}\left(  g^{\mu\nu}\phi_{\mu}\phi_{\nu}\right)  \right]  \ .
\end{gather}
Since $\mathcal{E}\left[  \phi\right]  $ is a total divergence, it easily
admits a first integral for static, spherically symmetric configurations.
\ Consider \emph{only} those situations in the following.

For such configurations the metric in generalized Schwarzschild coordinates is
\cite{Tolman}%
\begin{equation}
\left(  ds\right)  ^{2}=e^{N\left(  r\right)  }\left(  dt\right)
^{2}-e^{L\left(  r\right)  }\left(  dr\right)  ^{2}-r^{2}\left(
d\theta\right)  ^{2}-r^{2}\sin^{2}\theta\left(  d\varphi\right)  ^{2}\ .
\end{equation}
Thus for static, spherically symmetric $\phi$, with covariantly conserved
energy-momentum tensor (\ref{EnMomTensor}), Einstein's equations reduce to
just a pair of coupled $1$st-order nonlinear equations:%
\begin{align}
r^{2}\Theta_{t}^{\ t}  &  =e^{-L}\left(  rL^{\prime}-1\right)
+1\ ,\label{Albert1}\\
r^{2}\Theta_{r}^{\ r}  &  =e^{-L}\left(  -rN^{\prime}-1\right)  +1\ .
\label{Albert2}%
\end{align}
These are to be combined with the first integral of the $\phi$ field equation
in this situation. \ Defining
\begin{equation}
\eta\left(  r\right)  \equiv e^{-L\left(  r\right)  /2}\ ,\ \ \ \varpi\left(
r\right)  \equiv\eta\left(  r\right)  \phi^{\prime}\left(  r\right)  \ ,
\label{TwoDefns}%
\end{equation}
that first integral becomes
\begin{equation}
\frac{Ce^{-N/2}}{r^{2}}=\varpi\left(  1+\varpi^{2}\right)  +\frac{1}{2}\left(
N^{\prime}+\frac{4}{r}\right)  \eta\varpi^{2}\ , \label{1stIntegral}%
\end{equation}
where for asymptotically flat spacetime the constant $C$ is given by
$\lim_{r\rightarrow\infty}r^{2}\phi^{\prime}\left(  r\right)  =C$. \ Then upon
using
\begin{align}
\Theta_{t}^{\ t}  &  =\Theta_{\theta}^{\ \theta}=\Theta_{\varphi}^{\ \varphi
}=\tfrac{1}{2}\varpi^{2}\left(  1+\tfrac{1}{2}\varpi^{2}\right)  -\eta
\varpi^{2}\varpi^{\prime}\ ,\label{EM1}\\
\Theta_{r}^{\ r}  &  =-\tfrac{1}{2}\varpi^{2}\left(  1+\tfrac{3}{2}\varpi
^{2}\right)  -\tfrac{1}{2}\eta\varpi^{3}\left(  N^{\prime}+\tfrac{4}%
{r}\right)  \ , \label{EM2}%
\end{align}
the remaining steps to follow are clear.

First, for $C\neq0$, one can eliminate $N^{\prime}$ from (\ref{Albert2}) and
(\ref{1stIntegral}) to obtain an exact expression for $N$ in terms of $\eta$,
$\varpi$, and $C$:%
\begin{equation}
e^{N/2}=\frac{8C}{r\varpi}\frac{\eta-\frac{1}{2}r\varpi^{3}}{\left(
4\varpi-2r^{2}\varpi^{3}-r^{2}\varpi^{5}+8r\eta+12\varpi\eta^{2}+8r\varpi
^{2}\eta\right)  }\ . \label{ResultsOf1stIntegral}%
\end{equation}
If the numerator of this last expression vanishes there is an \emph{event
horizon}, otherwise not. \ When $\eta=\frac{1}{2}r\varpi^{3}$ the denominator
of (\ref{ResultsOf1stIntegral}) is positive definite. \ 

Next, in addition to (\ref{Albert1}) one can now eliminate $N$ from either
(\ref{Albert2}) or (\ref{1stIntegral}) to obtain two coupled first-order
nonlinear equations for $\eta$ and $\varpi$. \ These can be integrated, at
least numerically. \ Or they can be used to determine analytically the large
and small $r$ behaviors, hence to see if the energy and curvature are finite.
\ For example, again for asymptotically flat spacetime, it follows that%
\begin{align}
&  e^{L/2}\underset{r\rightarrow\infty}{\sim}1+\frac{M}{r}+\frac{1}{4}\left(
6M^{2}-C^{2}\right)  \frac{1}{r^{2}}+\frac{1}{2}M\left(  5M^{2}-2C^{2}\right)
\frac{1}{r^{3}}+O\left(  \frac{1}{r^{4}}\right)  \ ,\\
&  e^{N/2}\underset{r\rightarrow\infty}{\sim}1-\frac{M}{r}-\frac{1}{2}%
M^{2}\frac{1}{r^{2}}+\frac{1}{12}M\left(  C^{2}-6M^{2}\right)  \frac{1}{r^{3}%
}+O\left(  \frac{1}{r^{4}}\right)  \ ,\\
&  \varpi\underset{r\rightarrow\infty}{\sim}\frac{C}{r^{2}}\left(  1+\frac
{M}{r}+\frac{3}{2}M^{2}\frac{1}{r^{2}}\right)  +O\left(  \frac{1}{r^{5}%
}\right)  \ ,
\end{align}
for constant $C$ and $M$.

As of this writing the details of the two remaining first-order ordinary
differential equations are not pretty, but the equations are numerically
tractable. \ In terms of the variables defined in (\ref{TwoDefns}), Einstein's
equation (\ref{Albert1})\ becomes%
\begin{gather}
I\left(  r,\varpi,\eta\right)  r\frac{d}{dr}\varpi+J\left(  r,\varpi
,\eta\right)  r\frac{d}{dr}\eta=K\left(  r,\varpi,\eta\right)
\ ,\label{Poly1}\\
I\left(  r,\varpi,\eta\right)  =r\eta\varpi^{2}\ ,\ \ \ J\left(  r,\varpi
,\eta\right)  =-2\eta\ ,\\
K\left(  r,\varpi,\eta\right)  =\tfrac{1}{2}r^{2}\varpi^{2}\left(  1+\tfrac
{1}{2}\varpi^{2}\right)  +\eta^{2}-1\ .
\end{gather}
But worse than that, in light of (\ref{ResultsOf1stIntegral}) Einstein's
equation (\ref{Albert2}) becomes%
\begin{equation}
F\left(  r,\varpi,\eta\right)  r\frac{d}{dr}\varpi+G\left(  r,\varpi
,\eta\right)  r\frac{d}{dr}\eta=H\left(  r,\varpi,\eta\right)  \ ,
\label{Poly2}%
\end{equation}%
\begin{align}
F\left(  r,\varpi,\eta\right)   &  =-4\eta\left[  2r^{3}\varpi^{6}%
+3r^{3}\varpi^{8}+16\varpi\eta+4r\varpi^{4}\right. \nonumber\\
&  \left.  +16r\eta^{2}+48\varpi\eta^{3}+48r\varpi^{2}\eta^{2}+12r\varpi
^{4}\eta^{2}-12r^{2}\varpi^{5}\eta\right]  \ ,\\
G\left(  r,\varpi,\eta\right)   &  =8\eta\varpi^{2}\left[  2r^{2}\varpi
^{2}+3r^{2}\varpi^{4}-12\eta^{2}+12r\varpi^{3}\eta+4\right]  \ ,\\
H\left(  r,\varpi,\eta\right)   &  =\varpi\left[  8\eta\varpi\left(
4r\varpi^{3}-4\eta+2r^{2}\varpi^{2}\eta+3r^{2}\varpi^{4}\eta+12r\varpi^{3}%
\eta^{2}-12\eta^{3}\right)  \right. \nonumber\\
&  \left.  +\left(  4+3r^{2}\varpi^{4}+2r^{2}\varpi^{2}+12\eta^{2}\right)
\left(  4\varpi-r^{2}\varpi^{5}-2r^{2}\varpi^{3}+8r\varpi^{2}\eta
+8r\eta+12\varpi\eta^{2}\right)  \right]  .
\end{align}

As a representative example with $\varpi>0$, (\ref{Poly1}) and (\ref{Poly2})
were integrated numerically to obtain the results shown in Figure 1, for data
initialized as$\ \left.  \varpi\right\vert _{r=1}=0.5$ and $\left.
\eta\right\vert _{r=1}=1$. \ Evidently it is true that $\eta\left(  r\right)
\neq\frac{1}{2}r\varpi^{3}\left(  r\right)  $ for this case, so $e^{N\left(
r\right)  }$ does not vanish for any $r>0$ and there is no event horizon. \ 

However, there is a geometric singularity at $r=0$ with divergent scalar
curvature: $\ \lim_{r\rightarrow0}r^{3/2}R=const$. \ Since $R=-\Theta_{\mu
}^{\ \mu}$, and $\lim_{r\rightarrow0}\varpi$ is finite, this divergence in $R$
comes from the last term in (\ref{EM2}), which in turn comes from the second
term in $A$, i.e. the covariant $\partial\phi\partial\phi\partial^{2}\phi$ in
(\ref{Action}). \ In fact, it it not difficult to establish analytically for a
class of solutions of the model, for which the example in Figure 1 is
representative, the following limiting behavior holds.%
\begin{equation}
\lim_{r\rightarrow0}\left(  e^{L/2}/\sqrt{r}\right)  =\ell\ ,\ \ \ \lim
_{r\rightarrow0}\left(  \sqrt{r}e^{N/2}\right)  =n\ ,\ \ \ \lim_{r\rightarrow
0}\varpi=p\ ,\ \ \ \lim_{r\rightarrow0}\left(  \phi^{\prime}/\sqrt{r}\right)
=p\ell\ ,
\end{equation}
where $\ell$, $n$, and $p$ are constants related to the constant $C$ in
(\ref{1stIntegral}):%
\begin{equation}
2C=3np^{2}/\ell\ .
\end{equation}
It follows that for solutions in this class,%
\begin{equation}
\lim_{r\rightarrow0}r^{3/2}R=pC/n\ .
\end{equation}
For the example shown in Figure 1: \ $\ell\approx1.5$, $n\approx0.086$,
$p\approx3.3$, $C\approx0.94$, and $pC/n\approx36$.

For the same $\left.  \eta\right\vert _{r=1}=1$, further numerical results
show there are also curvature singularities without horizons for smaller
$\left.  \varpi\right\vert _{r=1}>0$, but event horizons are present for
larger scalar fields (roughly when $\left.  \varpi\right\vert _{r=1}>2/3$).
\ A more precise and complete characterization of the data set $\left\{
\left.  \varpi\right\vert _{r=1},\left.  \eta\right\vert _{r=1}\right\}  $ for
which there are naked singularities is in progress, but it is already evident
from the preceding remarks that the set has nonzero measure. \ 

The energy contained in \emph{only} the scalar field in the curved spacetime
is given by%
\begin{gather}
E_{\text{Galileon}}=\int_{0}^{\infty}\mathcal{H}\left(  r\right)
dr=\int_{-\infty}^{\infty}e^{s}\mathcal{H}\left(  e^{s}\right)  ds\ ,\\
\mathcal{H}\left(  r\right)  \equiv4\pi r^{2}e^{L/2}e^{N/2}\Theta_{t}%
^{\ t}=2\pi e^{2s}e^{L/2}e^{N/2}\varpi^{2}\left(  s\right)  \left(
1+\tfrac{1}{2}\varpi^{2}\left(  s\right)  \right)  -4\pi e^{s}e^{N/2}%
\varpi^{2}\left(  s\right)  \tfrac{d}{ds}\varpi\left(  s\right)  \ .
\end{gather}
For the above numerical example, the integrand $e^{s}\mathcal{H}\left(
e^{s}\right)  $ is shown in Figure 2. \ Evidently, $E_{\text{Galileon}}$\ is
finite in this case. \ It is also clear from the Figures that the Galileon
field has significant effects on the geometry in the vicinity of the peak of
its radial energy density. \ There the metric coefficients are greatly
distorted from the familiar Schwarzschild values, and as a consequence, the
horizon is eliminated.

It remains to investigate the stability of the static solutions described
above, and to consider the dynamical evolution of generic Galileon and other
matter field initial data, along the lines of \cite{Choptuik},\ to determine
under what physical conditions the naked singularities discussed here are
actually formed.

\begin{acknowledgments}
\textit{I thank D Fairlie for many discussions about Galileons in general, and
this work in particular. \ I also thank S K Rama, N Rinaldi, R Wald, and K S
Virbhadra for discussions of naked singularities. \ This research was
supported by a University of Miami Cooper Fellowship, and by NSF Awards
PHY-0855386 and PHY-1214521.}
\end{acknowledgments}

%

\begin{center}
\includegraphics[
height=4.0133in,
width=6.0265in
]%
{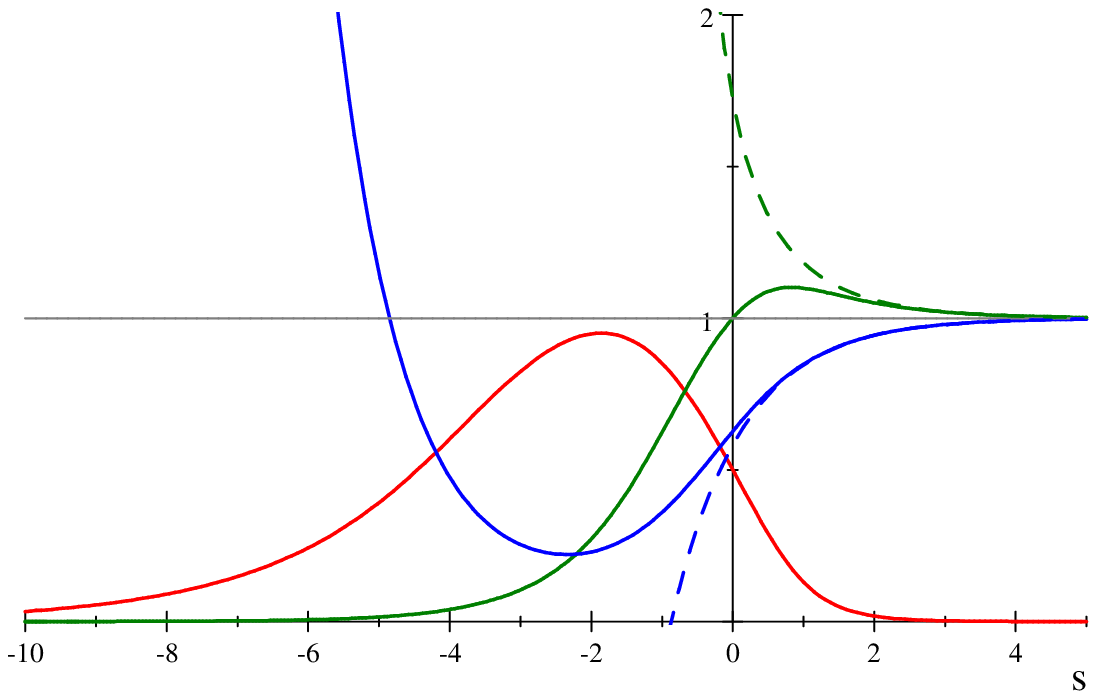}%
\\
Fig. 1: \ For initial values $\left.  {\protect\footnotesize \varpi
(s)}\right\vert _{s=0}=0.5$ and$\ \left.  {\protect\footnotesize \eta
(s)}\right\vert _{s=0}=1.0$, $d\phi/dr=\varpi/\eta$ is shown in red,
$e^{{\protect\footnotesize L}}=1/\eta^{{\protect\footnotesize 2}}$ in green,
and $e^{{\protect\footnotesize N}}$ in blue, where
$r=e^{{\protect\footnotesize s}}$. \ For comparison, Schwarzschild
$e^{{\protect\footnotesize L}}$ and $e^{{\protect\footnotesize N}}$ are also
shown as resp. green and blue dashed curves for the same $M\approx0.21$
\cite{Correction}.
\end{center}
%

\begin{center}
\includegraphics[
height=4.0136in,
width=6.0269in
]%
{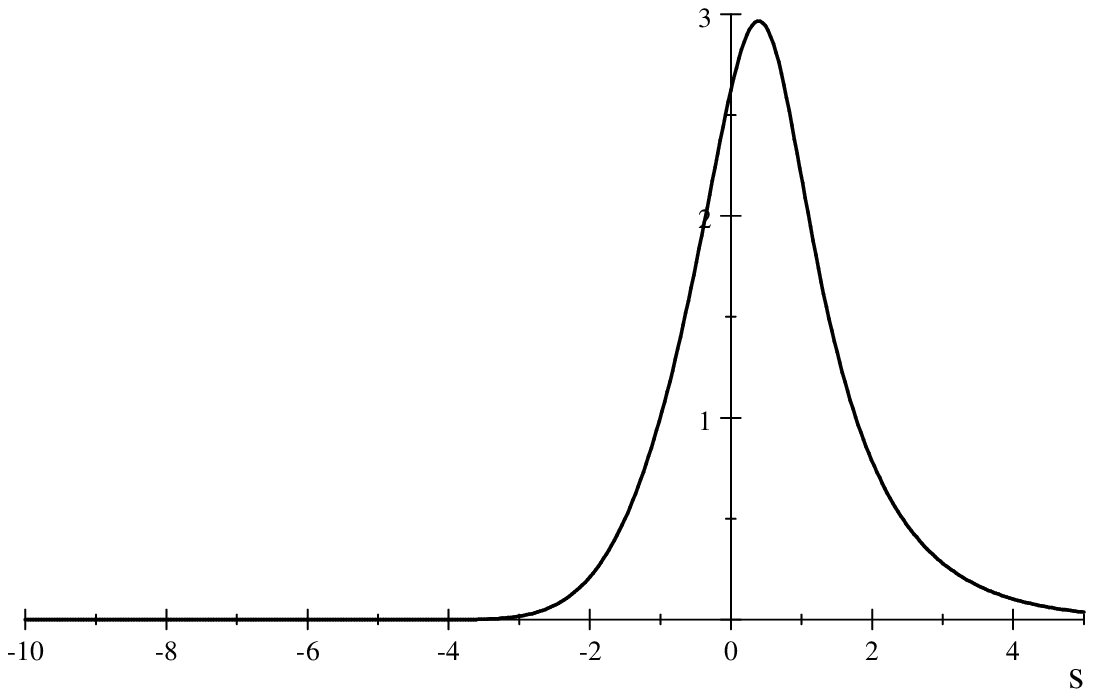}%
\\
Fig. 2: $\ e^{s}\mathcal{H}\left(  e^{s}\right)  $ for $\left.  \varpi\left(
s\right)  \right\vert _{s=0}=0.5$ and$\ \left.  \eta\left(  s\right)
\right\vert _{s=0}=1.0$, where $r=e^{{\protect\footnotesize s}}$.
\end{center}


\begin{thebibliography}{99}
\bibitem {Correction}Figure 1 in the original version of this manuscript
displayed the Schwarzschild metric for an erroneous mass (namely,
$M\approx0.36$, rather than the correct value that is used here,
$M\approx0.21$).
\end{thebibliography}
\end{document}